\documentclass[aps,prl,twocolumn,titlepage]{revtex4}

\usepackage{graphicx}
\usepackage{color}
\usepackage{dcolumn}
\usepackage{latexsym}
\usepackage[normalem]{ulem}
\usepackage{hyperref,amssymb}
\usepackage{url}
\usepackage{color,soul}
\usepackage{graphicx}
\usepackage{verbatim}
\usepackage{multirow}
\usepackage{amsmath}
\newcommand{\beq}{\begin{eqnarray}}
\newcommand{\eeq}{\end{eqnarray}}
\usepackage{mathrsfs}
\usepackage{float,soul}
\usepackage[dvipsnames]{xcolor}
\usepackage{mathtools}
\usepackage{slashed}
\usepackage{physics}	
\usepackage{graphicx}   
\usepackage{epstopdf}
\usepackage{subfigure}  
\usepackage{hyperref}   
\usepackage{bbold}
\usepackage{wasysym}
\usepackage{feynmp}
\usepackage{hyperref}
\hypersetup{colorlinks,
}

\usepackage[utf8]{inputenc}

\begin{document}

\title{\large Burgers rings as topological signatures of Eshelby-like plastic events in glasses}
\author{Arabinda Bera$^{1}$}
\author{Ido Regev$^{2}$}
\author{Alessio Zaccone$^{1}$}
\thanks{\color{blue}alessio.zaccone@unimi.it\color{black}}
\author{Matteo Baggioli$^{3}$}
\thanks{\color{blue}b.matteo@sjtu.edu.cn\color{black}}
\address{$^{1}$Department of Physics ``A. Pontremoli", University of Milan, via Celoria 16, 20133 Milan, Italy}
\address{$^{2}$ Environmental Physics, Jacob Blaustein Institutes for Desert Research, Ben-Gurion University of the Negev, Sede Boqer Campus 84990, Israel}
\address{$^3$Wilczek Quantum Center \& School of Physics and Astronomy, Shanghai Jiao Tong University, Shanghai 200240, China}

\begin{abstract}
Eshelby-like quadrupolar structures serve as the fundamental microscopic units for characterizing plastic instabilities in amorphous solids and play a crucial role in explaining their mechanical failure, including the formation of shear bands. However, identifying Eshelby-like plastic events in glasses remains challenging due to their inherent structural and dynamical complexity. In this work, we show that Eshelby-like structures can be precisely identified and localized using a topological invariant known as the continuous Burgers vector. By combining analytical and simulation techniques, we reveal the emergence of a topological Burgers ring around Eshelby plastic events, enabling the precise identification of their center of mass and capturing their orientation as well. This proposed method offers a clear and unambiguous framework to locate and characterize the plastic rearrangements that govern plasticity in glasses.
\end{abstract}

\maketitle
\color{blue}\textit{Introduction} \color{black} -- The description and quantification of plasticity and mechanical failure in crystalline materials \cite{schmid1968plasticity} are based on the concept of structural defects \cite{sutton2020physics}, particularly topological defects (TDs), such as dislocations. A typical example is the Peierls-Nabarro theory of yield stress \cite{Peierls_1940,Nabarro_1947}. In contrast, the elementary processes of plastic deformation in amorphous solids remain an open question \cite{annurev:/content/journals/10.1146/annurev-conmatphys-062910-140452,CRPHYS_2021__22_S3_117_0,berthier2024yieldingplasticityamorphoussolids}, with significant implications for their practical applications. 

This challenge primarily stems from the absence of an underlying lattice structure, which in crystals provides a well-defined ``background'' reference for identifying discrete defects. However, it was recognized early on that, in amorphous solids, structural relaxation tends to occur in localized regions exhibiting large non-affine displacements \cite{PhysRevE.57.7192} and local stress drops \cite{PhysRevE.82.066116}. These atomic rearrangements can be associated with quadrupolar zones and effectively described using the Eshelby inclusion model \cite{doi:10.1098/rspa.1957.0133,doi:10.1098/rspa.1959.0173}. 
These Eshelby-like plastic events have been incorporated into several mesoscopic models of plasticity in glasses, grounded in the successful yet phenomenological concept of ``soft spots'' or ``shear transformation zones (STZs) \cite{annurev:/content/journals/10.1146/annurev-conmatphys-062910-140452}. Moreover, the dynamics and, in particular, the alignment of these events have been identified as key factors in the formation of shear bands \cite{PhysRevLett.109.255502,PhysRevB.95.134111}, ultimately leading to plastic failure. Up to now, Eshelby-like quadrupolar defects and STZs remain the most widely used theoretical frameworks for characterizing and predicting plasticity in glasses, and are widely applied across a broad range of materials (e.g., \cite{PhysRevE.93.053002,PhysRevE.94.022907}).

Importantly, the stress drop induced by relaxation can be effectively characterized by the location and orientation of fictitious Eshelby inclusions \cite{PhysRevE.93.053002,PhysRevE.97.063002}. Nevertheless, beyond idealized and simplified scenarios, identifying the precise location and orientation of these Eshelby-like plastic events remains highly non-trivial. This detail is crucial for linking microscopic dynamics to macroscopic behavior.

As a concrete example, the spatial organization and sequence of STZ activations govern interaction pathways that underlie hysteresis and memory effects under cyclic loading \cite{RevModPhys.91.035002}, determine whether local rearrangements trigger avalanches or remain isolated, and provide essential input for mesoscale elastoplastic models of shear-band formation and failure \cite{RevModPhys.90.045006}. Previous efforts to identify STZ centers have relied on manual \cite{sastry2019} or semi-manual procedures \cite{https://doi.org/10.1063/5.0087164}, or on fitting quadrupolar displacement fields, approaches that quickly become unreliable when multiple events occur simultaneously. Locating STZ centers is therefore immediately valuable for testing theoretical predictions and advancing mesoscale modeling.

As comprehensively reviewed in \cite{RevModPhys.80.61}, well-defined topological defects, particularly dislocation-type defects, can also be defined in disordered condensed matter systems, including amorphous solids. In these systems, dislocation-like TDs are still characterized by a continuous topological invariant: the Burgers vector. However, unlike in crystals, this vector is continuous-valued, and no longer ``quantized" in units of the lattice spacing.

Building on these concepts, it was recently shown \cite{PhysRevLett.127.015501,Landry} that the continuous Burgers vector, computed from the non-affine displacement field, serves as a useful indicator of plasticity in glasses, accurately predicting the location of the yielding instability. Recent work \cite{liu2024measurablegeometricindicatorslocal} further suggests that this topological invariant provides a robust and measurable metric for characterizing local structural transformations associated with plastic events.

More broadly, these developments are part of a growing research program that reintroduces the concepts of topology and topological defects to describe and rationalize plasticity in amorphous solids \cite{PhysRevLett.127.015501,Baggioli2023,liu2024measurablegeometricindicatorslocal,wu2023topology,10.1093/pnasnexus/pgae315,Vaibhav2025,PhysRevB.110.014107,huang2024spottingstructuraldefectscrystals,bera2025hedgehogtopologicaldefects3d,wu2024geometrytopologicaldefectsglasses,PhysRevE.109.L053002,zheng2025topologicalsignaturescollectivedynamics}. Notably, Ref. \cite{PhysRevE.109.L053002} demonstrated that antivortex defects in the displacement field are strongly reminiscent of Eshelby inclusions in two-dimensional disordered solids, suggesting a possible method for identifying STZs. Additionally, topological defects have been used to reinterpret shear banding in metallic glasses \cite{PhysRevB.110.014107} and to rationalize their formation through the emergence and dynamics of vortices \cite{SOPU2023170585}.

In this work, we advance the idea that topological concepts play a crucial role in disordered systems such as glasses. Specifically, we show that the continuous Burgers vector can be effectively employed to identify Eshelby-like plastic events in two-dimensional systems. By combining analytical insights with simulation data, we demonstrate the emergence of Burgers ``rings'' encircling quadrupolar Eshelby-like structures (plastic events). Furthermore, these rings allow one to quantitatively determine the center of the Eshelby-type plastic event, something that could not be achieved thus far with traditional Eshelby or STZ-type analyses.

\begin{figure*}
  \centering
   \includegraphics[width=\linewidth]{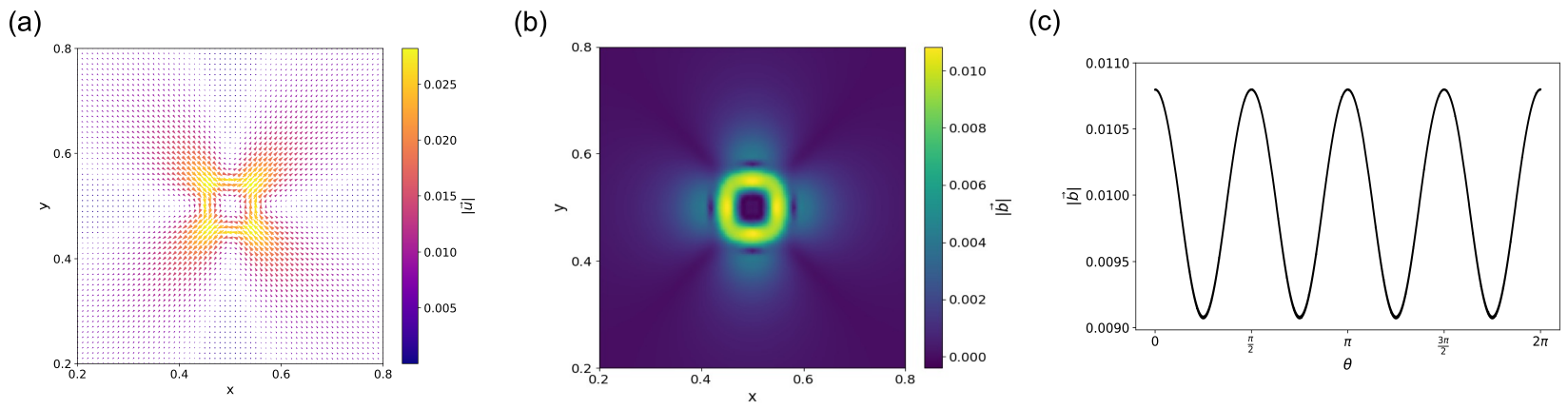}
    \caption{\textbf{(a)} The displacement field for an ideal Eshelby inclusion centered at $(x,y)=(0.5,0.5)$ is shown. The color bar indicates the magnitude of the displacment field. For this case we have fixed $\varepsilon^*=1$, $\nu=0.46$, $\phi=\pi/4$ and $a=0.05$. \textbf{(b)} The magnitude of the Burgers vector is shown with a color bar. The circular ring around $r=a$ is prominent.  \textbf{(c)} The variation of $|\vec{b}|$ along the azimuthal angle $\theta$ along the Burgers ring $r=a$.}
    \label{fig1}
\end{figure*}

\color{blue}\textit{Ideal Eshelby inclusion and Burgers ring} \color{black} -- We start by considering the idealized Eshelby inclusion problem \cite{doi:10.1098/rspa.1957.0133,doi:10.1098/rspa.1959.0173}. The analytical solution for the elastic displacement field $\vec{u}\equiv(u_x,u_y)$ of a 2D Eshelby inclusion is given by (see \cite{PhysRevE.87.022810} for a derivation):
\begin{widetext}
\begin{align}
   & u^{out}_x(r > a)=\frac{\varepsilon^*}{4(1-\nu)}\frac{a^2}{r^2}\Big{\{} \Big{[}  2(1-2\nu)+\frac{a^2}{r^2} \Big{]}[x\cos2\phi+y\sin2\phi]+  \Big{[}1-\frac{a^2}{r^2}\Big{]} \Big{[} \frac{(x^2-y^2)\cos2\phi+2xy\sin2\phi}{r^2} \Big{]}2x  \Big{\}},\nonumber\\
    & u^{out} _y(r > a)=\frac{\varepsilon^*}{4(1-\nu)}\frac{a^2}{r^2}\Big{\{} \Big{[}  2(1-2\nu)+\frac{a^2}{r^2} \Big{]}[x\sin2\phi-y\cos2\phi]+  \Big{[}1-\frac{a^2}{r^2}\Big{]} \Big{[} \frac{(x^2-y^2)\cos2\phi+2xy\sin2\phi}{r^2} \Big{]}2y  \Big{\}}, \label{th}
\end{align}
\end{widetext}
and:
\begin{align}
   & u^{in}_x(r < a)=\frac{\varepsilon^*}{4(1-\nu)}  (3-4\nu) [x\cos2\phi+y\sin2\phi],\nonumber\\
    & u^{in}_y(r < a)=\frac{\varepsilon^*}{4(1-\nu)}  (3-4\nu) [x\sin2\phi-y\cos2\phi], \label{in_filed}
\end{align}
where $out$-$in$ correspond respectively to the region outside ($r>a$) and inside ($r<a$) the inclusion, with $r\equiv \sqrt{x^2+y^2}$. We notice that the displacement field is continuous at $r=a$ but its derivatives with respect to $x,y$ are not. Here, $\varepsilon^*$ is the eigenstrain magnitude, $\nu$ is the Poisson ratio, and $a$ is the radius of the inclusion. The parameter $\phi$ represents the orientation of the quadrupolar zone, while $r$ denotes the radial distance from the quadrupole's center. The coordinates $(x, y)$ specify a point relative to the quadrupole's center, which is taken for simplicity as the origin $(0,0)$. The displacement field of an ideal Eshelby inclusion, corresponding to the expressions in Eqs.~\eqref{th}-\eqref{in_filed}, is shown in Fig.~\ref{fig1}(a). The displacement field is illustrated for an Eshelby inclusion centered at $(0.5, 0.5)$ with a radius $a = 0.05$. The other parameters are set to $\varepsilon^*=1$, $\nu=0.46$ and $\phi=\pi/4$. The characteristic quadrupolar symmetry is evident.

Leveraging on the analytical expression, Eq.~\eqref{th} and Eq.~\eqref{in_filed}, we compute the Burgers vector $\vec{b}$ \cite{kleinert1989gauge,PhysRevLett.127.015501},
\begin{equation}
    b_i=-\oint_{\mathcal{L}} du_i=-\oint_{\mathcal{L}}\label{def}
\frac{du_i}{dx^k}dx^k,
\end{equation}
with $\mathcal{L}$ a closed loop of arbitrary shape, that for simplicity is taken to be a circle of radius $R = a/2$ centered at $(x_0, y_0)$ (see Supplementary Material, SM). The Burgers vector is then evaluated by systematically varying the loop center $(x_0, y_0)$ over the two-dimensional grid. The Burgers vector magnitude for this ideal case is shown in Fig.~\ref{fig1}(b). The value of the Burgers vector is nonzero only along a localized ring, from now on labeled as the \textit{Burgers ring}, that coincides with the core $r=a$ of the (fictitious) inclusion. The Burgers ring arises due to the discontinuity of the strain tensor at the inclusion boundary $r=a$. Consequently, any loop integral defined as in Eq.~\eqref{def}, where the integration path $\mathcal{L}$ intersects the circle $r=a$, yields a finite value of $\vec{b}$.

Interestingly, the amplitude of $\vec{b}$ along the Burgers ring captures the orientation of the Eshelby inclusion, as demonstrated in Fig.~\ref{fig1}(c). Indeed, the amplitude of the Burgers vector along the ideal Burgers ring, $r=a$, shows a periodic pattern with minima at azimuthal angle $\theta=\pi/4,3 \pi/4, 5 \pi/4,7 \pi/4$. The first minimum of $|\vec{b}|$ aligns with the principal axis of the Eshelby inclusion in Fig.~\ref{fig1}(a), in this representative example.

In the \textit{End Matter} (see Appendix A), we show that this method also applies to an aligned array of Eshelby-like quadrupoles, a structure known to develop along shear bands in disordered solids under mechanical deformation \cite{PhysRevE.87.022810}.

{\color{blue}\textit{Topological defects and plastic spots}} -- In order to go beyond the ideal scenario discussed above, we simulate a two-dimensional glass model \cite{regev2013}. Detail of the system are provided in the Supplementary Material (SM). In Fig. \ref{fig2}(a) we show a particle displacement field generated by a plastic event (see Supplementary Material for further explanations). In this case, there is one  Eshelby-like ``soft-spot'' \cite{manning2011vibrational}, featuring approximate quadrupolar symmetry, which is marked with a green circle for better clarity.

We interpolate the displacement field onto a $32 \times 32$ lattice grid. In Fig. \ref{fig2}(b) we show the normalized displacement field $\hat{u}\equiv \vec{u}/|\vec{u}|$, mapped onto the same lattice grid, where the corresponding color indicates its amplitude $|\vec{u}|$. After normalizing the vector field, spatial structures (e.g., swirlings) appear. In order to describe the topology of the displacement field, we resort to the concept of winding number $q$. In two dimensions, the local winding number is given by:
\begin{equation}
    q=\frac{1}{2 \pi} \oint_{\mathcal{L}} d\theta,
\end{equation}
where $\theta$ is the direction of $\vec{u}$. When $\mathcal{L}$ is chosen as the smallest loop possible, then local regions with $q=\pm 1$ correspond respectively to the presence of a vortex/anti-vortex topological defect.

These vortex-like topological defects have been recently related to the characteristics of Eshelby inclusions \cite{PhysRevE.109.L053002}. However, as we will explicitly demonstrate, the concept of vortex-like topological defects alone is insufficient to accurately identify quadrupolar Eshelby-like plastic events in two-dimensional amorphous systems. 

Red and blue circles in Fig.~\ref{fig2}(b) identify respectively vortices ($q=+1$) and anti-vortices ($q=-1$). Consistent with the analysis of \cite{PhysRevE.109.L053002}, we find that an anti-vortex emerges at the core of the Eshelby-like structure. Nevertheless, it is not the only one: we observe other topological defects with a rather ordered spatial correlation. We notice that the total winding number within the simulation box remains zero due to the application of periodic boundary conditions in both directions and the Poincar\'e-Hopf index theorem \cite{DoCarmo}.

\begin{figure}[ht]
    \centering
    \includegraphics[width=0.8\linewidth]{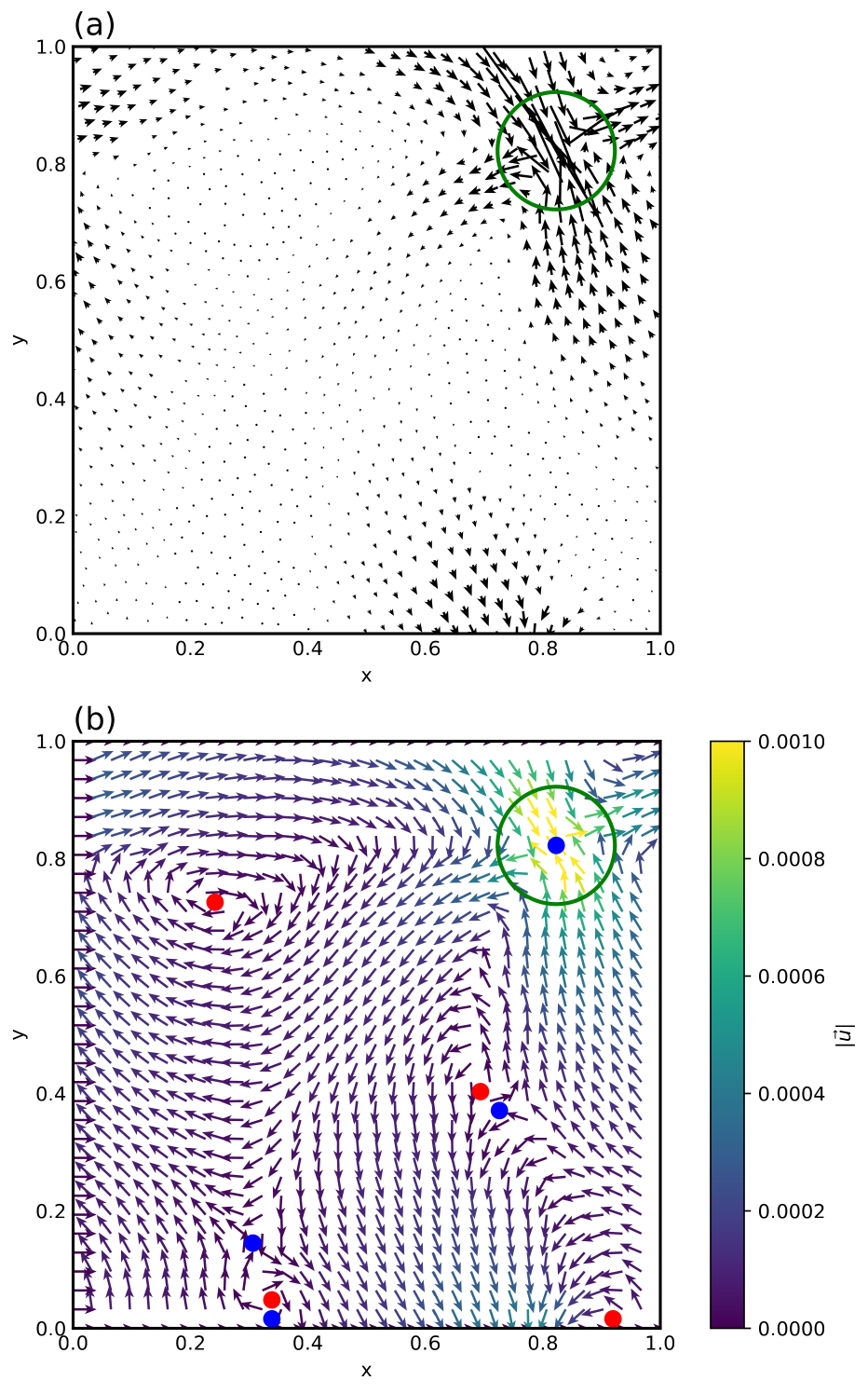}
    \caption{\textbf{(a)} Particle displacement field. The circle highlights the region where an Eshelby-like structure emerges. \textbf{(b)} Normalized displacement vector field on a $32 \times 32$ lattice grid. Topological defects with winding numbers $+1$ and $-1$ are marked in red and blue, respectively. The color bar indicates the magnitude of the displacement field.}
    \label{fig2}
\end{figure}

\begin{figure}[ht]
    \centering
    \includegraphics[width=\linewidth]{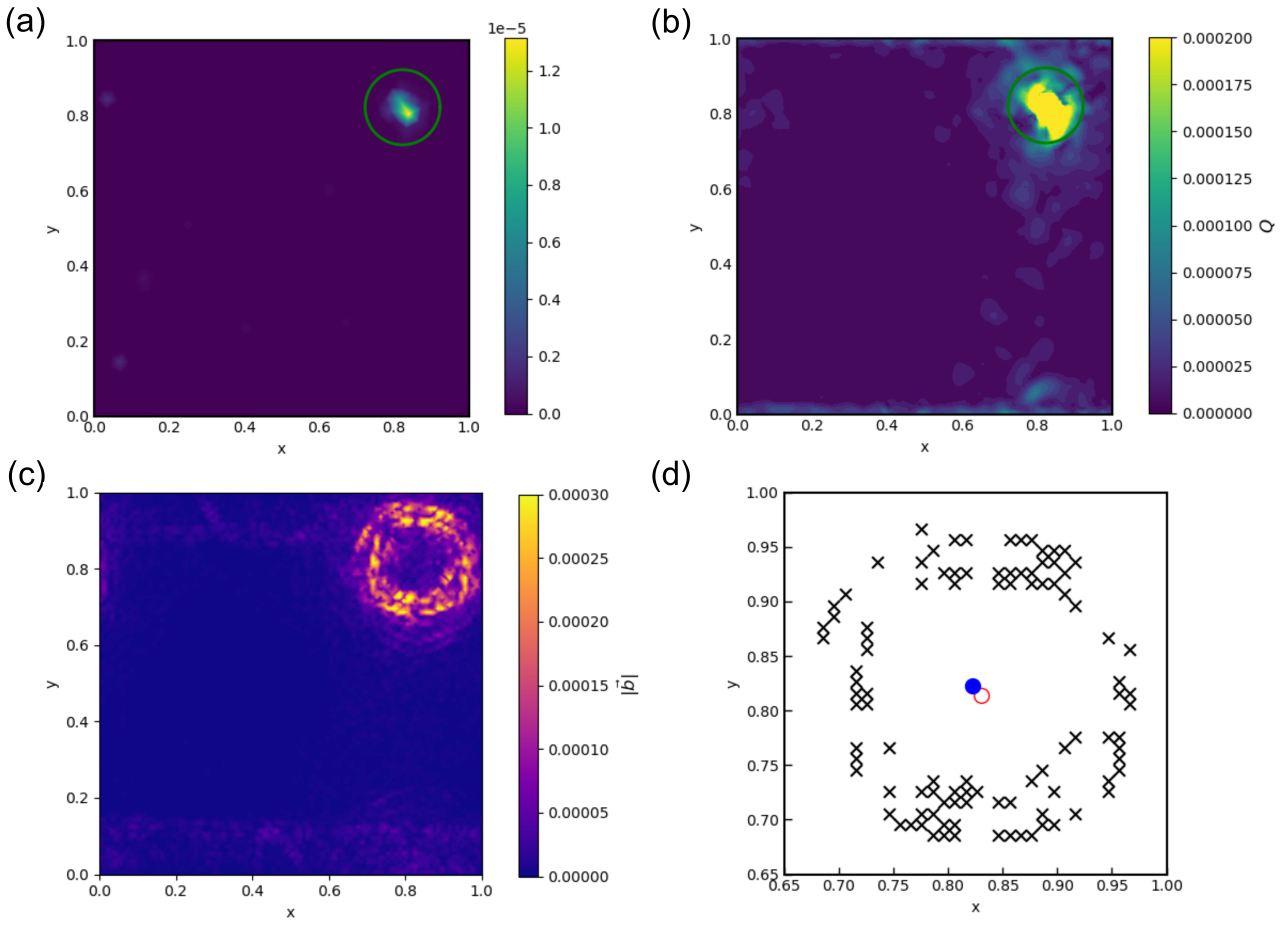}
    \caption{\textbf{(a)} Spatial maps of the non-affine displacement measure $D^2_{\rm{min}}$. \textbf{(b)} The magnitude of quadrupole charge $Q$ at lattice grid is shown. The color bar represents the value of $Q$. The highlighted circle marks the region associated with an Eshelby-like rearrangement. \textbf{(c)} The magnitude of the Burgers vector is shown. \textbf{(d)} The locations having highest value of $|\vec{b}|$ are represented by black crosses. Their resulting center of mass is indicated with an open red circle. The location of the $-1$ TD is indicated by a blue filled circle. }
    \label{fig3}
\end{figure}

In order to properly identify the plastic spots, we further look for the relative particle displacement with respect to their neighbors and compute the standard metric $D^2_{\text{min}}$ \cite{PhysRevE.57.7192} (see details in SM). As shown in Fig.~\ref{fig3}(a), the magnitude of $D^2_{\text{min}}$ is highest around the Eshelby-like structure and confirms that not all topological antivortex-like defects in Fig.~\ref{fig2}(b) are related to plastic spots. 

Moreover, in Fig.~\ref{fig3}(b) we show the amplitude of the quadrupolar charge $Q$ (see for example \cite{PhysRevE.108.L042901}), defined by decomposing the symmetric strain tensor $\mathbf{s}$ as
\begin{equation}
\mathbf{s} = m\mathbf{I} + \mathbf{Q},
\end{equation}
with $\mathbf{I}$ being the two-dimensional identity matrix. The quadrupole tensor $\mathbf{Q}$ can be expressed in its standard form as \cite{PhysRevE.92.062403}
\begin{equation}
\mathbf{Q}=Q
\begin{bmatrix}
\cos(2\vartheta) & \sin(2\vartheta) \\
\sin(2\vartheta) & -\cos(2\vartheta)\\
\end{bmatrix},
\end{equation}
where the magnitude is defined as $Q = \frac{1}{2}\sqrt{(s_{xx} - s_{yy})^2 + 4 s_{xy}^2}$, and the orientation as $\vartheta =(1/2)\arctan\ \left[{2s_{xy}}/{(s_{xx}-s_{yy})}\right]$, with $s_{ij} = \frac{1}{2}(\partial_i u_j + \partial_j u_i)$ for $(i,j) = (x,y)$. See the SM for further details.

As shown in Fig.~\ref{fig3}(b), the Eshelby-like plastic event in Fig.~\ref{fig2}(a) has a strong quadrupolar nature, corresponding to a large value of $Q$. On the other hand, in the rest of the $(x,y)$ plane, $Q$ vanishes or it is at least negligibly small. 

Several key conclusions can be drawn from the investigation of the simulation data. (I) The positions of (anti)vortex-like topological defects alone are insufficient to precisely identify the Eshelby-like structures associated with plastic events. (II) The distribution of $D^2_{\text{min}}$ and $Q$ offers only an approximate indication of the location of these events and does not allow for an accurate determination of their center of mass. (III) There is no evident correlation between the locations of (anti)vortex-like topological defects in the displacement field and either $D^2_{\text{min}}$ or $Q$, casting doubt on the effectiveness of (anti)vortex-like defects as indicators of plasticity.

{\color{blue}\textit{Burgers ring}} -- We then compute the local value of the Burgers vector $\vec{b}$ on the simulated 2D displacement field using a closed circular loop of radius $R=0.12$, while varying the loop center ($x_0,y_0$) across the simulation box, similar to the ideal Eshelby case. The results corresponding to the case presented in Fig.~\ref{fig2}(a) are shown in Fig.~\ref{fig3}(c). We observe that, like the results in the idealized case in Fig.~\ref{fig1}(b), the Burgers vector is generally zero everywhere apart from a thin shell that is located exactly around the Eshelby-like quadrupolar structure. This result suggests that, even in a more realistic scenario, Eshelby-like plastic events are surrounded by a Burgers ring. Nevertheless, within the precision of our numerical data we are not able to ascertain whether the distribution of the Burgers vector captures the orientation of the Eshelby-like structure as well, as previously demonstrated for the idealized case. Partially, this might be also caused by the fact that in the simulations the Eshelby-like events do not display perfect quadrupolar symmetry and therefore their orientation is not neatly defined. 

Three additional cases are presented in Appendix B in the \textit{End Matter}, confirming the universality of our findings. Moreover, the stability of the results upon changing the loop radius $R$ is demonstrated in the SM.

To further demonstrate the ability of the Burgers ring to locate the plastic events, in Fig. \ref{fig3}(d) we present a zoomed-in view where the top $1\%$ of highest values of $|\vec{b}|$ around the plastic event are marked with black crosses. As also evident from this representation, a circular ring is formed around the Eshelby-like structure. We then compute the center of mass of these points, that is indicated with a red open circle. We find that the center of mass, identified using the Burgers ring method, is very close to the $-1$ topological defect lying at the center of the quadrupolar Eshelby-like structure, and already presented in Fig.~\ref{fig2}(b).

\begin{figure}[ht]
    \centering
    \includegraphics[width=1\linewidth]{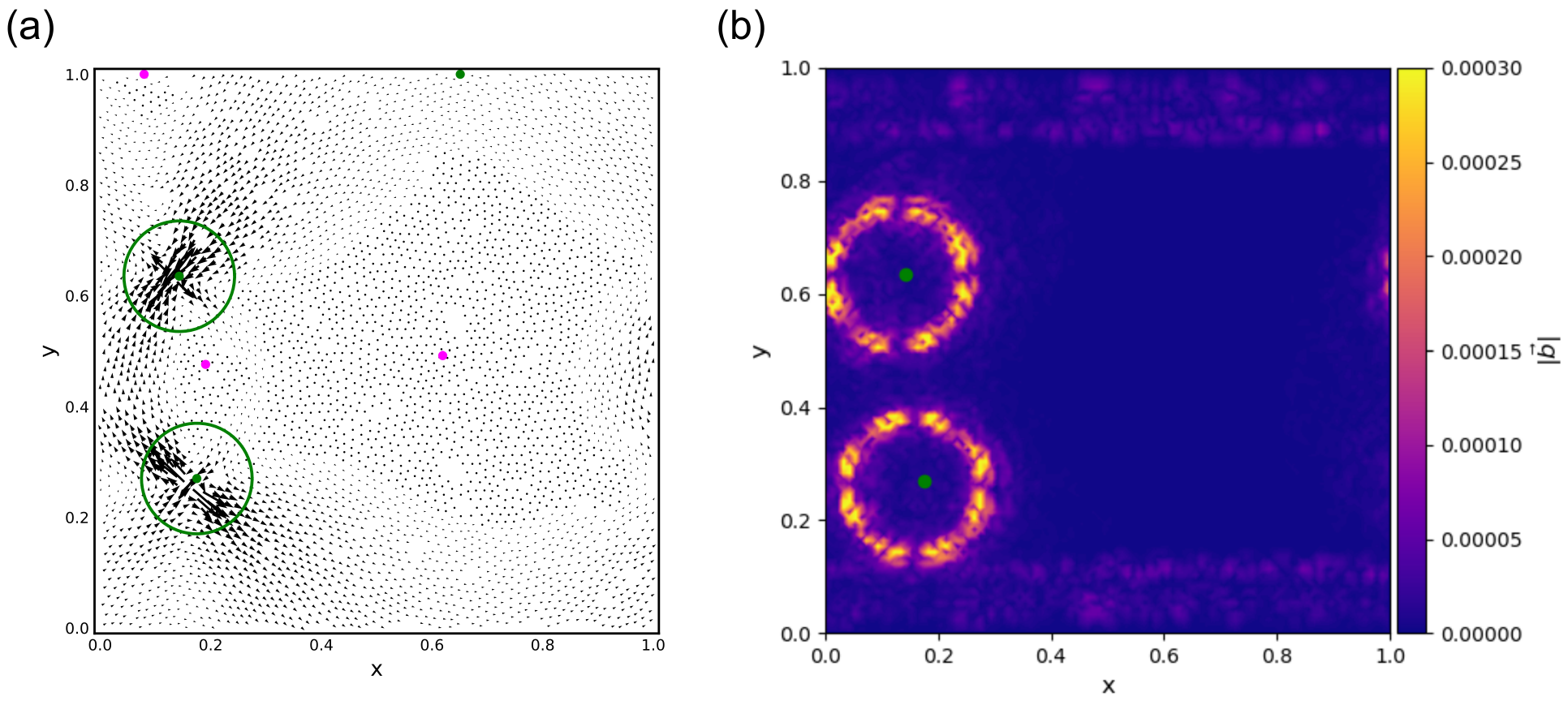}
    \caption{\textbf{(a)} Snapshot of the particle displacement field for a configuration with $N=4096$. Two different Eshelby-like events are shown by green circles. The topological defects with $q=+1$ and $q=-1$ are shown in magenta and green filled circles. \textbf{(b)} Continuous Burgers vector field computed with loop radius $R=0.12$. Two Burgers rings emerge at the location of the Eshelby-like events. The green filled circles indicate the location of the $-1$ topological defects shown also in panel (a) and roughly identifying the center of the Eshelby-like events.}
    \label{fig4}
\end{figure}

We further explore the applicability of our method to configurations with multiple Eshelby events. In Fig.~\ref{fig4}(a), we display the particle displacement field for a system containing two well-separated Eshelby events, whose centers are highlighted by green circles that coincide with $q=-1$ topological defects. In Fig.~\ref{fig4}(b), the corresponding Burgers vector field is computed, revealing two distinct Burgers rings encircling the Eshelby centers. In addition, vortex-like defects with $q=+1$ emerge at the interface between the two events as shown in Fig. \ref{fig4}(a) (also see Fig. \ref{fig_em2} in the End Matter), closely resembling the multiple Eshelby configurations predicted in idealized analytical scenarios. Further examples are presented in Fig.~\ref{fig_em3}, Appendix~C in the \textit{End Matter}. When the events are well separated, distinct Burgers rings appear, whereas for nearby inclusions the rings overlap and only partial structures are observed. In summary, these results demonstrate that the proposed method is both robust and effective in capturing multiple, interacting Eshelby events.

\color{blue}\textit{Conclusions} \color{black} -- In this work, we have shown that the local value of the continuous Burgers vector $\vec{b}$ can be effectively used to identify Eshelby-like quadrupolar plastic events in two-dimensional amorphous solids. Specifically, $\vec{b}$ becomes nonzero only along a localized structure, a ``Burgers ring'', that encircles the core of the Eshelby-like deformation.

Compared to other existing methods, such as (anti)vortex-like topological defects \cite{wu2023topology}, the $D^2_{\text{min}}$ metric \cite{PhysRevE.57.7192}, and the quadrupolar charge $Q$ \cite{PhysRevE.108.L042901}, the Burgers ring approach offers significant advantages. In particular, (anti)vortex-like topological defects fail to reliably identify the location of Eshelby-like events, as they also appear in regions unrelated to plastic deformation. Additionally, unlike $D^2_{\text{min}}$ or the quadrupolar charge $Q$, our method enables precise identification of the center of mass of these quadrupolar plastic events, in good agreement with the location of the antivortex defect \cite{PhysRevE.109.L053002}.

Beyond single events, we further demonstrated that our method remains robust in configurations involving multiple, interacting Eshelby inclusions. Even when Burgers rings overlap or become distorted, the event centers can still be approximately determined from the partial ring by treating it as a segment of a circular structure. This highlights the robustness of the approach in complex configurations. This ability to robustly locate and characterize individual shear transformation zones provides crucial microscopic input for mesoscale elastoplastic models, and directly connects to the mechanisms underlying hysteresis, memory formation, avalanche statistics, and shear-band evolution \cite{RevModPhys.91.035002,RevModPhys.90.045006,sastry2019,https://doi.org/10.1063/5.0087164}.

In summary, we provide a robust and well-defined topological metric to accurately identify and locate Eshelby-like plastic events in 2D glasses. In the future, it would be interesting to extend this program to particle-resolved experimental systems such as colloidal glasses, foams, and granular films, as already attempted in \cite{liu2024measurablegeometricindicatorslocal}, and to three-dimensional systems, where other topological measures have been already applied \cite{Cao2018,bera2025hedgehogtopologicaldefects3d,wu2024geometrytopologicaldefectsglasses}. 

\color{blue}{\it Acknowledgments} \color{black} -- AZ is indebted to Prof. Srikanth Sastry for stimulating discussions and for providing a motivation to carry out this research. MB acknowledges the support of the Shanghai Municipal Science and Technology Major Project (Grant No.2019SHZDZX01) and the support of the sponsorship from the Yangyang Development Fund. AZ gratefully acknowledges funding from the European Union through Horizon Europe ERC Grant number: 101043968 ``Multimech''. AB and AZ gratefully acknowledge funding from the US Army DEVCOM Army Research Office through Contract No. W911NF-22-2-0256. IR was supported by the Israel Science Foundation (grant No. 1204/23).

\section*{End Matter}\label{end}

{\it Appendix A: Multiple ideal Eshelby events-} We further test the method in more complex scenarios involving multiple interacting Eshelby inclusions, following the methods in \cite{PhysRevE.87.022810}. For a system of $N_{\rm esb}$ inclusions, the displacement field follows a linear superposition of the single inclusion solutions, ${\bf u}_k^{\rm out}(x_k,y_k;a_k,\varepsilon_k^*,\nu,\phi_k)$ and ${\bf u}_k^{\rm in}(x_k,y_k;a_k,\varepsilon_k^*,\nu,\phi_k)$, where $(x_k,y_k)$ denotes the local coordinates relative to the center of the $k^{\rm th}$ inclusion of radius $a_k$. The explicit forms are given in Eqs.~\eqref{th} and \eqref{in_filed}. Outside all inclusion cores, the displacement field can be written as \cite{PhysRevE.87.022810}
\begin{equation}\label{eq4}
{\bf u}^{\rm out}(x,y)=\sum_{k=1}^{N_{\rm esb}} {\bf u}_k^{\rm out}(x_k,y_k;a_k,\varepsilon_k^*,\nu,\phi_k),
\end{equation}
while inside the $m^{\rm th}$ inclusion one obtains the inside field:
\begin{equation}\label{eq5}
\begin{aligned}
{\bf u}^{\rm in,m}(x,y) &= 
   {\bf u}_m^{\rm in}(x_m,y_m;a_m,\varepsilon_m^*,\nu,\phi_m) \\
   &\quad + \sum_{k\neq m}{\bf u}_k^{\rm out}(x_k,y_k;a_k,\varepsilon_k^*,\nu,\phi_k).
\end{aligned}
\end{equation}
For a single inclusion the strain inside the core is strictly uniform. In contrast, for multiple inclusions the local strain within inclusion $m$ acquires additional contributions from the nonuniform fields generated by all other inclusions.

\begin{figure}
\centering
\includegraphics[width=\linewidth]{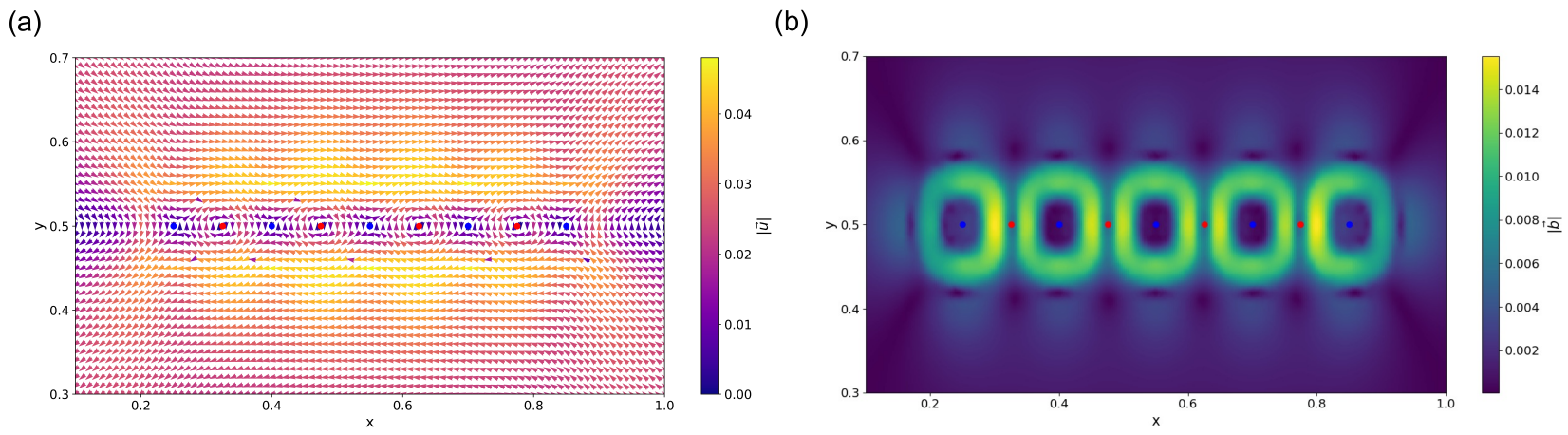}
\caption{\textbf{(a)} Normalized displacement field from the analytical solution for five Eshelby inclusions aligned along a $x$-axis at $x=0.25, 0.4, 0.55, 0.70$ and $0.85$. Each inclusion uses the same parameters as in the single inclusion case of Fig.~\ref{fig1}\textbf{(a)}. Blue and red symbols correspond respectively to the topological defects with negative and positive winding number. The color bar represents the  magnitude of the resultant field. \textbf{(b)} The heat map of the amplitude of the continuous Burgers vector field for the same configuration in \textbf{(a)}. Five aligned Burgers rings emerge at the location of the Eshelby-like plastic events.}
\label{fig_em2}
\end{figure}

In Fig.~\ref{fig_em2}(a) we show the normalized displacement field obtained from Eqs.~\eqref{eq4} and \eqref{eq5} for $N_{\rm esb}=5$ inclusions aligned along the $x$-axis, with the color bar indicating the field magnitude. Eshelby centers are marked by blue circles, while red circles denote vortex-like defects that appear along the interfaces connecting consecutive inclusions. To test the Burgers-vector method, we compute the continuous Burgers vector using Eq.~\eqref{def} and plot its magnitude in Fig.~\ref{fig_em2}(b). Five distinct Burgers rings are observed, each centered on an Eshelby core. When inclusions are placed in close proximity, the rings become distorted and appear only partially.\color{black}

\begin{figure}
\centering
\includegraphics[width=\linewidth]{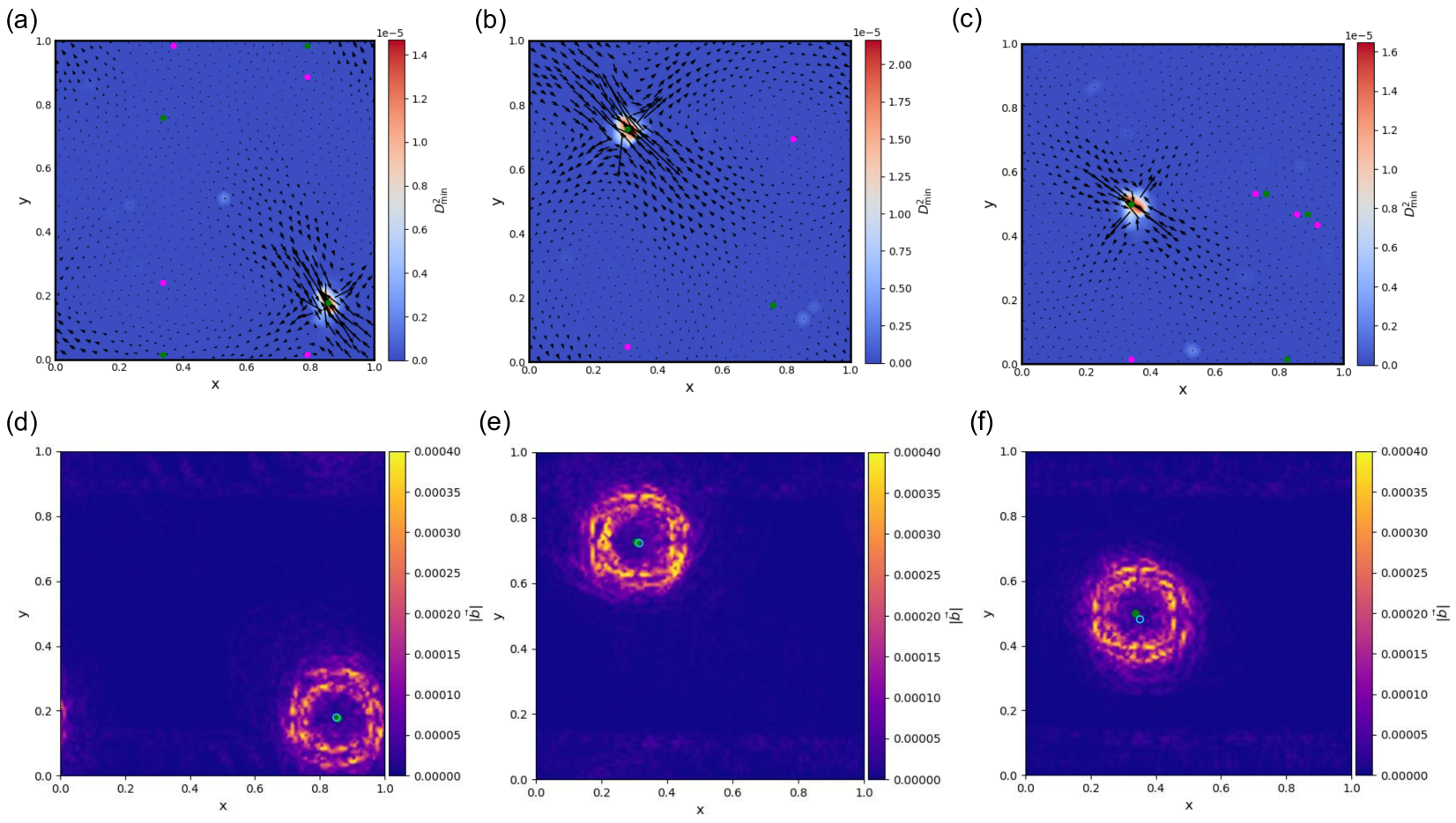}
\caption{\textbf{(a-c)} Several snapshots of the displacement vector field during plastic events. The background color map indicates the local value of $D^2_{\text{min}}$ and helps identifying the quasi-localised Eshelby-like soft spots. The positions of topological defects with winding numbers $q=+1$ and $-1$ are indicated by magenta and green filled circles, respectively. \textbf{(d-f)} The corresponding map of the Burgers vector amplitude $|\vec{b}|$. The Burgers rings surrounding the Eshelby-like plastic events are evident. The white open circle is the center of mass of the Burgers ring, while the green disk is the location of the $-1$ anti-vortex topological defect located at the center of the Eshelby-like plastic spot.}
\label{fig_em}
\end{figure}

{\it Appendix B: Universality of our findings -} In order to confirm the universality of our findings, we have studied the displacement field and its characteristics across various plastic events observed in the stress-strain curve. In Fig.~\ref{fig_em}, we present three representative cases. Panels (a-c) show the displacement vector field (black arrows) with a superimposed color map of the coarse-grained $D^2_{\text{min}}$. In all three cases, isolated Eshelby-like plastic spots are recognized and correspond to regions with large $D^2_{\text{min}}$. In the corresponding panels (d-f), we display the local value of the Burgers vector amplitude $|\vec{b}|$. The locations of topological defects with winding number $-1$ inside the Burgers rings and the center of mass of the top $1\%$ highest $|\vec{b}|$ values are shown by green filled circles and an open circle, respectively.
As evident, in all cases we observed a very neat Burgers ring surrounding the Eshelby-like plastic events and confirming the findings presented in the main text. \\

\begin{figure}
\centering
\includegraphics[width=\linewidth]{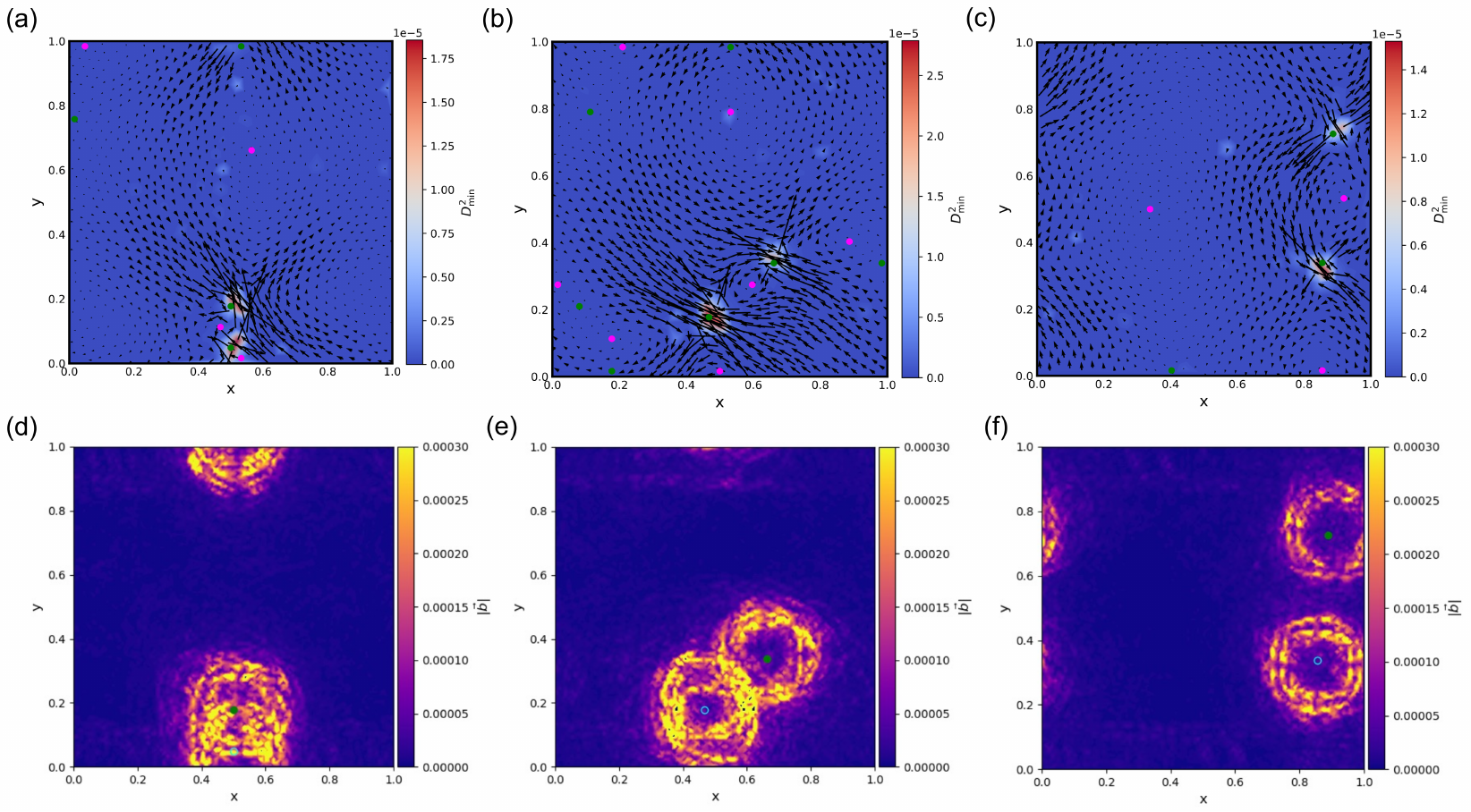}
\caption{\textbf{(a–c)} Snapshots of the displacement vector field containing two Eshelby events at different locations. The background color shows the local value of $D^2_{\text{min}}$, highlighting quasi-localized Eshelby-like soft spots. Magenta and green filled circles mark topological defects with winding numbers $q=+1$ and $q=-1$, respectively. \textbf{(d–f)} Corresponding Burgers vector amplitude $|\vec{b}|$, revealing the superposition of two Burgers rings around the Eshelby-like plastic events. The two distinct $q=-1$ antivortex defects at the centers of the rings are indicated by open and filled circles.}
\label{fig_em3}
\end{figure}

{\it Appendix C: Simulated configurations with multiple Eshelby-like events-} We further present displacement fields for simulated configurations exhibiting two Eshelby events in Fig.~\ref{fig_em3}(a–c). The background color shows the coarse-grained $D^2_{\text{min}}$, while magenta and green circles denote topological defects with winding numbers $q=+1$ and $q=-1$, respectively. Figures~\ref{fig_em3}(d–f) display the corresponding Burgers vector fields, with the $q=-1$ defects inside the Burgers rings marked by open and filled circles. When the two events are close, the Burgers rings overlap and appear as partial circular structures, whereas for well-separated events, two distinct rings are clearly observed.

\newpage
\onecolumngrid
\appendix 
\clearpage
\renewcommand\thefigure{S\arabic{figure}}    
\setcounter{figure}{0} 
\renewcommand{\theequation}{S\arabic{equation}}
\setcounter{equation}{0}
\renewcommand{\thesubsection}{SM\arabic{subsection}}
\section*{\Large Supplementary Material}
\section*{Simulation details and stress-strain curve}
We consider a two-dimensional system of point-like particles interacting via a radially-symmetric, attractive-repulsive, pairwise potential $U(r)$, where $r$ denotes the distance between particle centers. 
The pair-wise potential consists of the repulsive part of the standard Lennard-Jones potential, connected via a hump
to a region that is smoothed continuously to zero.
Specifically, the interaction potential takes the form \cite{lerner2009locality,regev2013}:

\begin{equation}\label{eq:full_potential}
U(r) =
\begin{cases}
\varepsilon \left[ \left( \dfrac{\sigma}{r} \right)^{12} - \left( \dfrac{\sigma}{r} \right)^6 + \dfrac{1}{4} - h_0 \right], & r \leq \sigma x_0, \\
\varepsilon h_0 \, P\left( \dfrac{r/\sigma - x_0}{x_c} \right), & \sigma x_0 < r \leq \sigma (x_0 + x_c), \\
0, & r > \sigma(x_0 + x_c).
\end{cases}
\end{equation}

Here, $\varepsilon$ sets the energy scale, and $\sigma$ is the characteristic length scale of interaction. The potential minimum occurs at $x_0 = 2^{1/6}$, corresponding to the standard location of the minimum in the Lennard-Jones potential. Beyond this minimum, the potential is smoothly deformed using a polynomial $P(x)$, ensuring continuity up to the second derivative and allowing a gradual decay to zero at $r = \sigma(x_0 + x_c)$. The parameter $h_0$ controls the depth of the potential well. All of the variables and constants are given in Lennard-Jones units \cite{regev2013,allen2017computer}. 

The smoothing function $P(x)$ is a sixth-order polynomial defined as:
\begin{equation}\label{eq:poly}
P(x) = \sum_{i=0}^{6} A_i x^i,
\end{equation}
with coefficients:
\[
\begin{aligned}
&A_0 = -1.0, \quad A_1 = 0.0, \quad A_2 = 1.785826183464224, \\
&A_3 = 28.757894970278530, \quad A_4 = -81.988642011620980, \\
&A_5 = 76.560294378549440, \quad A_6 = -24.115373520671220.
\end{aligned}
\]
The variation of  $U(r)$ with $r$ is shown in Supplementary Fig.~\ref{fig_S1}(a). 
Our simulation included a system of $N$ particles interacting via the potential $U(r)$ where for each pair of particles $i$ and $j$ the value of $\sigma$ was chosen to be:
\[
\sigma = \frac{\sigma_i + \sigma_j}{2},
\]
where $\sigma_i$ is the radius of particle $i$ and $\sigma_j$ is the radius of particle $j$. To prevent crystallization, we used a binary mixture where half the particles had a radius $\sigma_i=1$ and the other half had a radius of $\sigma=1.4$. The system was kept at a constant volume with a constant density of $\rho=0.75$, in which the system is in a jammed state. In the figures shown in the text, the coordinates are normalized such that the simulation square has a side of size $1$. All results are shown for $N=1024$, unless stated otherwise.

To prepare amorphous solids we first simulated a liquid at high temperature, using leap-frog integration combined with the Berendsen thermostat. We then cooled the liquid to a lower temperature and minimized the energy to zero using the FIRE minimization algorithm \cite{bitzek2006structural}. The protocol details are given in \cite{regev2013}.

A representative snapshot of the resulting amorphous solid is shown in Supplementary Fig.~\ref{fig_S1}(b), where small black circles indicate particles of radius $1$ and large red circles represent particles of size $1.4$. We then applied shear using the athermal quasistatic (AQS) scheme where shear strain steps of magnitude $10^{-4}$ using the Lees-Edwards periodic boundary conditions \cite{lees1972computer} (black arrows in Supplementary Fig.~\ref{fig_S1}(b) indicate  one of two possible shear directions). After each strain step, the energy was minimized using the FIRE algorithm in order to achieve a mechanically stable, zero temperature, configuration. When repeated many times, this scheme produces a typical stress-strain curve, as is shown in Supplementary Fig.~\ref{fig_S1}(c). In this curve, straight lines indicate elastic response, while stress drops indicate plastic events. The Eshelby structures discussed in the text are all results of such plastic transitions.

\begin{figure}[h]
\centering
\includegraphics[width=\linewidth]{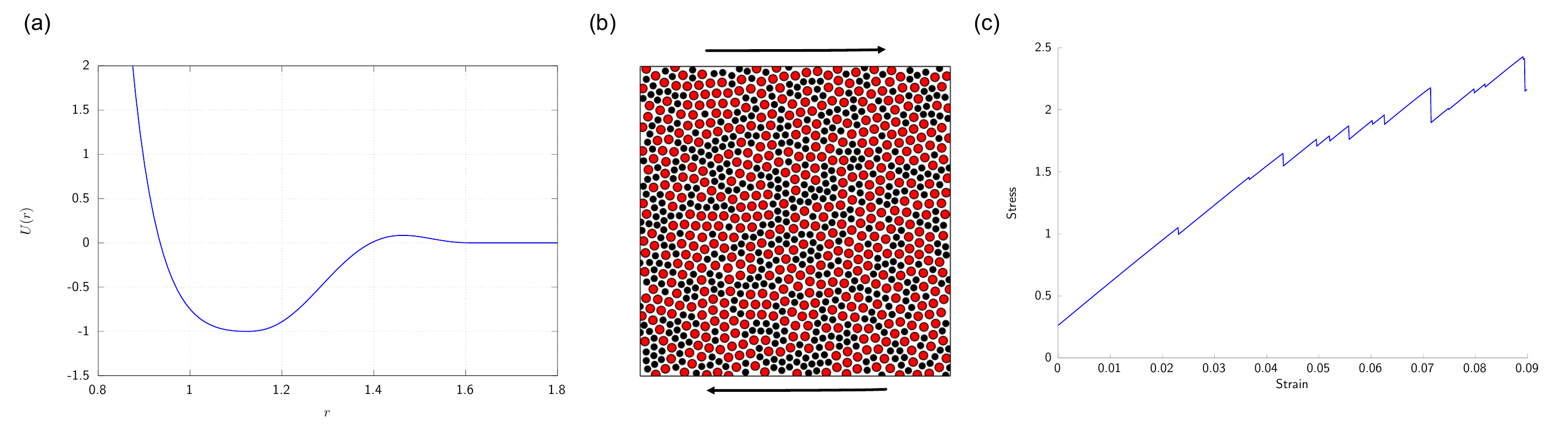}
\caption{\textbf{(a)} The interaction potential, Eq. \ref{eq:full_potential} with $h_0=1$, $\epsilon=1$, $\sigma=1$ and $x_c=0.5$. \textbf{(b)} A typical simulation box with arrows showing the directions of shear. \textbf{(c)} A typical stress-strain curve generated from the AQS simulation. We can observe that the increase in the stress as a function of strain is linear most of the time, but that once in a while there is a sharp drop in the stress, indicating a plastic event similar to the ones discussed in the  text.}
\label{fig_S1}
\end{figure}
\section*{Interpolation of the displacement field}
The particle displacement field obtained from simulations is first interpolated onto a regular grid to identify the topological defects. We employ the standard Python routine ``\textit{griddata}'' to interpolate the displacement field $\vec{u}=(u_x,u_y)$ onto a square lattice of size $32\times32$ for $N=1024$ and $64\times64$ for $N=4096$. This step provides a smooth and continuous representation of the displacement field across the simulation domain.

\section*{Analytical and numerical computation of the Burgers vector}
\color{black}In a two-dimensional space parameterized by Cartesian coordinates $(x,y)$, the Burgers vector can be written as:
\begin{equation}\label{eq10}
b_x =-\oint_{\mathcal{L}} \left( \frac{\partial u_x}{\partial x}dx+\frac{\partial u_x}{\partial y}dy \right),\qquad b_y =-\oint_{\mathcal{L}} \left( \frac{\partial u_y}{\partial x}dx+\frac{\partial u_y}{\partial y}dy \right),
\end{equation}
with ${\mathcal{L}}$ being a closed loop of arbitrary shape.

For simplicity, we consider a circular loop of radius $R$ centered at $(x_0,y_0)$, described in parametric form by $x=x_0+R\cos (t)$ and $y=y_0+R\sin (t)$, where the variable $t$ varies from $0$ to $2\pi$. Substituting these expressions, we obtain:
\begin{equation}\label{eq12}
b_x =\int_0^{2\pi} dt \, R \left(\frac{\partial u_x}{\partial x}\sin t - \frac{\partial u_x}{\partial y}\cos t \right),\qquad b_y =\int_0^{2\pi} dt \, R \left(\frac{\partial u_y}{\partial x}\sin t - \frac{\partial u_y}{\partial y}\cos t \right).
\end{equation}
Now the Burgers vector is computed with with fixed $R$ by varying the center of the loop ($x_0,y_0$) throughout the simulation box. 

To compute the Burgers vector $\vec{b}$ numerically in simulation data, we evaluate the line integral of the displacement gradient around a closed loop of fixed radius $R$ centered at each grid point $(x_0, y_0)$. In practice, the derivatives of the interpolated displacement field, $\partial u_x / \partial x$, $\partial u_x / \partial y$, $\partial u_y / \partial x$, and $\partial u_y / \partial y$, are computed using central finite differences. For improved accuracy, the derivatives are further smoothed using the Python spline interpolation routine ``\textit{RectBivariateSpline}''. The line integral along the circular path is then approximated numerically using the trapezoidal rule over $1000$ discretized points. This procedure is repeated for all grid points to construct a map of the Burgers vector field across the system. The resulting $|\vec{b}|$ field captures the presence of Eshelby-like events, including overlapping or partially formed Burgers rings that appear when multiple events occur in close proximity.

\color{black}To test the robustness of our findings, we present the spatial distribution of the Burgers vector magnitude $|\vec{b}|$ for three different loop radii $R$ in Supplementary Fig.~\ref{fig_S2}(a–c). The appearance of the Burgers ring is evident in all cases. To assess the predictive capability of the Eshelby center, we identify the top $1\%$ of the highest $|\vec{b}|$ values and show that their center of mass approximately coincides with the position of the topological defects with winding number $-1$ in that region, as illustrated in Supplementary Fig.~\ref{fig_S2}(d–f).

\begin{figure}[h]
\centering
\includegraphics[width=\linewidth]{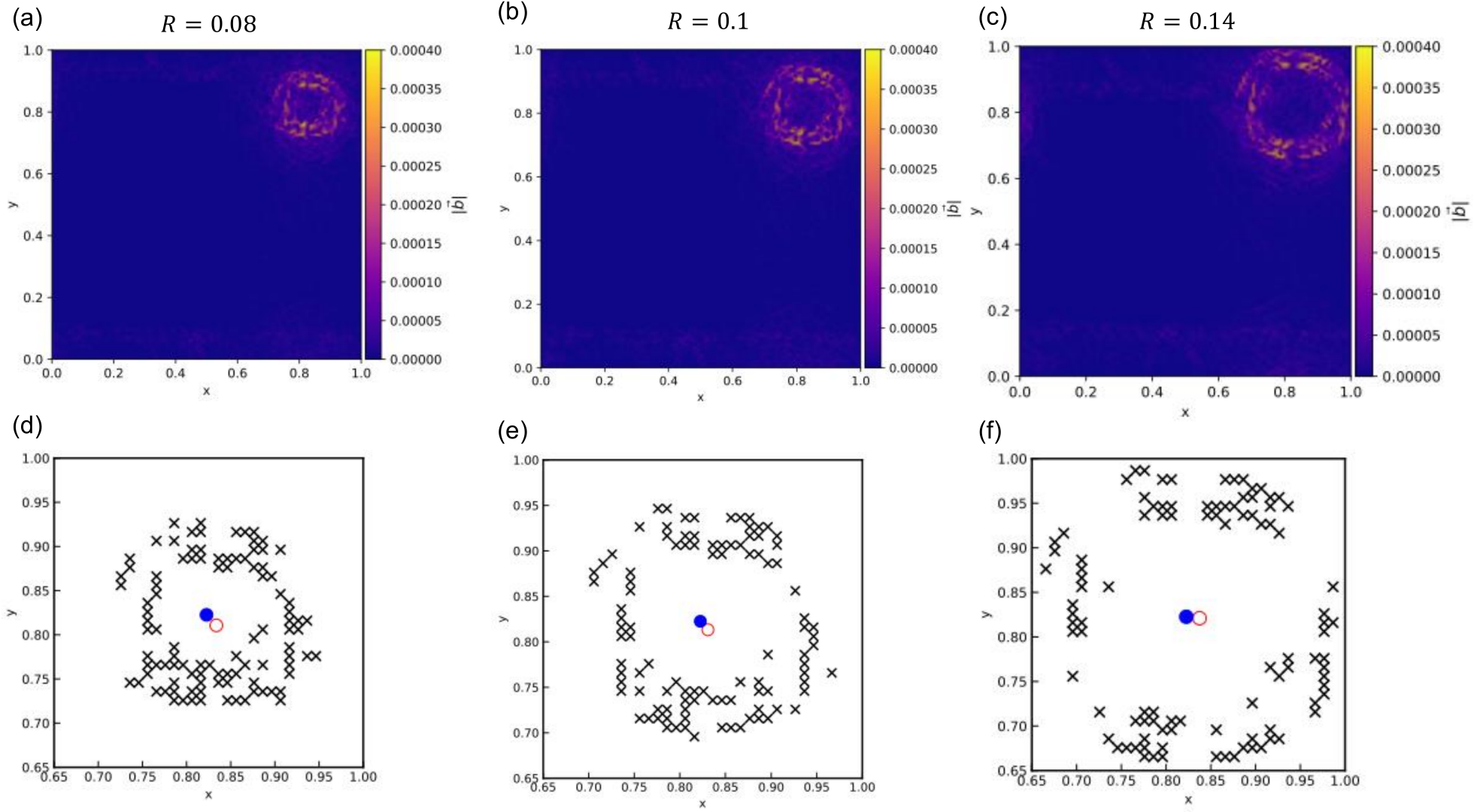}
\caption{\textbf{(a–c)} Magnitude of the Burgers vector $|\vec{b}|$ computed using three different closed-loop radii, $ R = 0.08$, $0.01$, and $0.14$, respectively. The color bar indicates the value of $|\vec{b}|$. \textbf{(d–f)} Positions of the top $1\%$ highest $|\vec{b}|$ values are marked with black crosses. The center of mass of these points is indicated by red open circles, while the locations of topological defects with winding number $-1$ in the corresponding regions are shown as blue filled circles, for the data presented in panels (a), (b), and (c), respectively.}
\label{fig_S2}
\end{figure}

\section*{Calculation of $D^2_{\text{min}}$}
During athermal quasi-static (AQS) deformation, we track the particle positions at two successive strain steps, $\vec{r}(\gamma-\Delta \gamma)$ and $\vec{r}(\gamma)$. To quantify the local non-affine motion of particle $i$, we first identify its neighboring particles within a suitable cutoff $r_c$. The non-affine measure for particle $i$ is then defined as
\begin{equation}
D^2_i = \sum_{j=1}^{N_i} \left| \vec{r}_j(\gamma) - \vec{r}_i(\gamma) - {\bf F}_i \cdot \left[ \vec{r}_j(\gamma-\Delta\gamma) - \vec{r}_i(\gamma-\Delta\gamma) \right] \right|^2,
\end{equation}
where $j$ runs over all $N_i$ neighbors of particle $i$, and ${\bf F}_i$ is the best-fit local deformation tensor that minimizes $D^2_i$ \cite{PhysRevE.57.7192}. The resulting minimized value is denoted as $D^2_{\text{min}}$, which provides a measure of the extent to which particle motion deviates from an affine deformation in the local neighborhood. 
\color{black}
\begin{figure}[h]
\centering
\includegraphics[width=\linewidth]{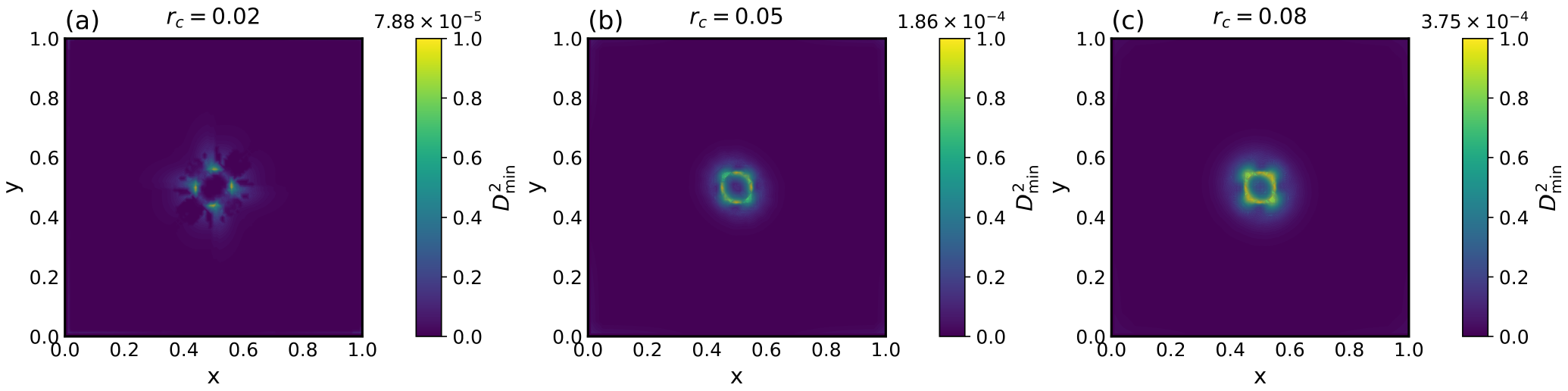}
\caption{\textbf{(a-c)} The magnitude of $D^2_{min}$ for an ideal Eshelby inclusion centered at $(x,y)=(0.5,0.5)$ for different interaction cut-off distance $r_c$. The parameters for the inclusion have been fixed to $a = 0.05$, $\varepsilon^*=1$, $\nu=0.46$ and $\phi=\pi/4$.}
\label{fig_S3}
\end{figure}

In order to compute $D^2_{\text{min}}$ for an Eshelby inclusion, we numerically construct a square grid of spacing $l_s=0.01$ within a unit-length square box. Then we displace each grid point according to the Eqs. 1-2 of the main article, considering them as point particles. Then we compute $D^2_{\text{min}}$ for these two consecutive configurations. In Supplementary Fig.~\ref{fig_S3} we show the magnitude of $D^2_{\text{min}}$ for different values of the interaction radius $r_c$ for an ideal Eshelby inclusion centered at $(x,y)=(0.5,0.5)$. The emergence of circular patterns surrounding the Eshelby center is clearly observable even in $D^2_{\text{min}}$ values. For the calculation of $D^2_{\text{min}}$ in the simulation data presented in the main text, we use a cutoff distance of $r_c=0.05$, which corresponds approximately to the first minimum of the pair correlation function $g(r)$ for the 2D glass sample. However, in simulation data, this quantity does not reflect the symmetric patterns observed in the ideal case, although it effectively captures the approximate regions of plastic deformation.  

\section*{Analytical solutions for the quadrupole tensor}
The quadrupole tensor is constructed from the displacement field $\vec{u}$. The strain tensor is first computed as
\begin{equation}
s_{ij} = \tfrac{1}{2}\left(\partial_i u_j + \partial_j u_i\right), \qquad (i,j)=(x,y),
\end{equation}
with explicit components in two dimensions given by
\begin{equation}
s_{xx} = \frac{\partial u_x}{\partial x}, \quad
s_{yy} = \frac{\partial u_y}{\partial y}, \quad
s_{xy} = s_{yx} = \tfrac{1}{2}\left( \frac{\partial u_x}{\partial y} + \frac{\partial u_y}{\partial x} \right).
\end{equation}
\color{black}
\begin{figure}
\centering
\includegraphics[width=0.35\linewidth]{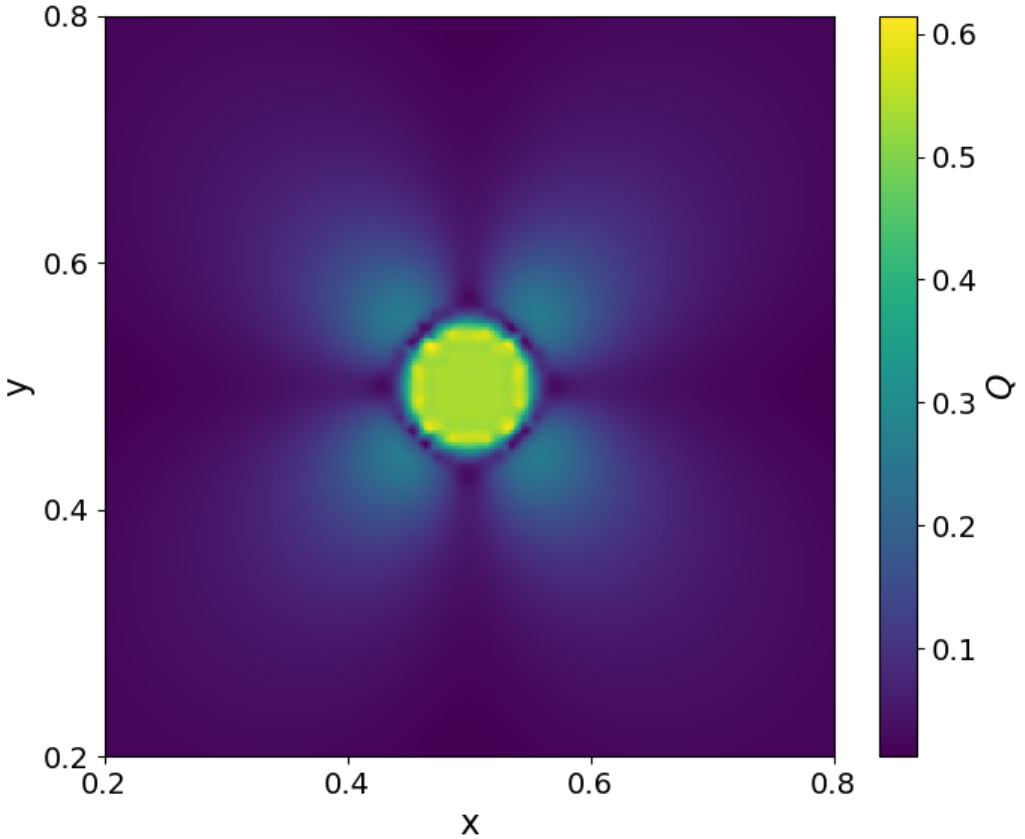}
\caption{The magnitude of the quadrupole charge $Q$, as defined in Eq. \eqref{mag}, for an ideal Eshelby inclusion centered at $(x_0,y_0)=(0.5,0.5)$. Parameters are set to $a = 0.05$, $\varepsilon^*=1$, $\nu=0.46$ and $\phi=\pi/4$.}
\label{fig_S4}
\end{figure} 
The strain tensor can be decomposed into an isotropic and a traceless part:
The latter can be decomposed as  
\begin{equation}
\begin{bmatrix}
s_{xx} & s_{xy} \\
s_{yx} & s_{yy}
\end{bmatrix}
=
\tfrac{1}{2}(s_{xx}+s_{yy})
\begin{bmatrix}
1 & 0 \\
0 & 1
\end{bmatrix}
+
\begin{bmatrix}
\tfrac{1}{2}(s_{xx}-s_{yy}) & s_{xy} \\
s_{xy} & -\tfrac{1}{2}(s_{xx}-s_{yy})
\end{bmatrix},
\label{eq:decomp}
\end{equation}
where the first term corresponds to the isotropic compression/dilation \(m\mathbf{I}\) with  
\[
m=\tfrac{1}{2}(s_{xx}+s_{yy}),
\]  
and the second term defines the quadrupole tensor \(\mathbf{Q}\).  

The traceless tensor \(\mathbf{Q}\) can be expressed in the standard form
\begin{equation}
\mathbf{Q}=Q
\begin{bmatrix}
\cos(2\vartheta) & \sin(2\vartheta) \\
\sin(2\vartheta) & -\cos(2\vartheta)
\end{bmatrix},
\end{equation}
with
\[
Q\cos(2\vartheta)=\tfrac{1}{2}(s_{xx}-s_{yy}), \qquad Q\sin(2\vartheta)=s_{xy},
\]
noting that \(s_{xy}=s_{yx}\) because the strain tensor is symmetric in the two indices.  

Hence, the quadrupole magnitude and orientation follow as
\begin{align}
Q &= \tfrac{1}{2}\sqrt{(s_{xx}-s_{yy})^2 + 4s_{xy}^2}, \label{mag}\\
\vartheta &= \tfrac{1}{2}\arctan\!\left(\frac{2s_{xy}}{s_{xx}-s_{yy}}\right).
\end{align}

\color{black}We use the analytical expressions for $s_{ij}$ in the case of an ideal Eshelby inclusion to evaluate the quadrupole magnitude $Q$. Supplementary Fig.~\ref{fig_S4} shows $Q$ for an inclusion centered at $(x_0,y_0)=(0.5,0.5)$ with parameters $a = 0.05$, $\varepsilon^*=1$, $\nu=0.46$, and $\phi=\pi/4$. The quadrupolar symmetry of the inclusion is clearly evident.

\end{document}